\documentclass[twocolumn,amsmath,amssymb,pre]{revtex4}

\usepackage[]{graphicx} 
\usepackage{amsmath}

\begin{document}

\title{A self-consistent field theory of density correlations in classical fluids}

\author{Hiroshi Frusawa}
\email{frusawa.hiroshi@kochi-tech.ac.jp}

\affiliation{Laboratory of Statistical Physics, Kochi University of Technology, Tosa-Yamada, Kochi 782-8502, Japan.}

\date{\today}

\begin{abstract}
More than half of a century has passed since the free energy of classical fluids defined by second Legendre transform was derived as a functional of density-density correlation function.
It is now becoming an increasingly significant issue to develop the correlation functional theory that encompasses the liquid state theory, especially for glassy systems where out of equilibrium correlation fields are to be investigated.
Here we have formulated a field theoretic perturbation theory that incorporates two-body fields (both of density-density correlation field and its dual field playing the role of two-body interaction potential) into a density functional integral representation of the Helmholtz free energy.
Quadratic density fluctuations are only considered in the saddle-point approximation of two-body fields as well as density field.
We have obtained a set of self-consistent field equations with respect to these fields, which simply reads a modified mean-field equation of density field where the bare interaction potential in the thermal energy unit is replaced by minus the direct correlation function given in the mean spherical approximation.
Such replacement of the interaction potential in the mean-field equation belongs to the same category as the local molecular field theory proposed by Weeks and co-workers in a series of papers [e.g., J. M. Rodgers  {\itshape et al.}, Phys. Rev. Lett., {\bf 97}, 097801 (2006); R. C. Remsing {\itshape et al.}, P. Natl. Acad. Sci. USA, {\bf 113}, 2819 (2016)].
Notably, it has been shown that even the mean-field part of the free energy functional given by the self-consistent field theory includes information on short-range correlations between fluid particles, similarly to the formulation of the local molecular field theory.
The advantage of our field theoretic approach is not only that the modified mean-field equation can be improved systematically, but also that fluctuations of two-body fields in nonuniform fluids may be considered, which would be relevant especially for glass-forming liquids where heterogeneous out-of-equilibrium correlations are to be investigated.
\end{abstract}


\maketitle

\section{Introduction}
The liquid state theory (LST) has investigated constituent particle arrangement, or particle-particle correlations, on a molecular scale [1, 2].
As a consequence, various approximate forms of correlation functions have been obtained from a couple of equations between the direct correlation function $c(r)$ and the total correlation function $h(r)$ (or the radial distribution function $g(r)\equiv h(r)-1$) that depend on the separation distance $r$ between particles:
the Ornstein-Zernike (OZ) equation and an approximate form of the closure relation such as the mean spherical approximation (MSA) and hypernetted chain (HNC) closures [1-10].
A great deal of knowledge on correlation functions, gained by the LST, has been incorporated into field theories, or coarse-grained theories described by fields as collective variables [11-80].

\begin{table*}
\caption{Comparison between a variety of field theories that consists of the functional theories of density and correlation fields [11-20], the variational theories based on the upper and lower bounds [1, 21-33], and the functional integral forms of one-body potential and density fields [34-56]. To clarify the relationship between the field theories and the LST, second and third column describes the uses of density-density correlation functions and reference fluids from the LST, respectively.}
\begin{center}
\includegraphics[width=11.5cm]{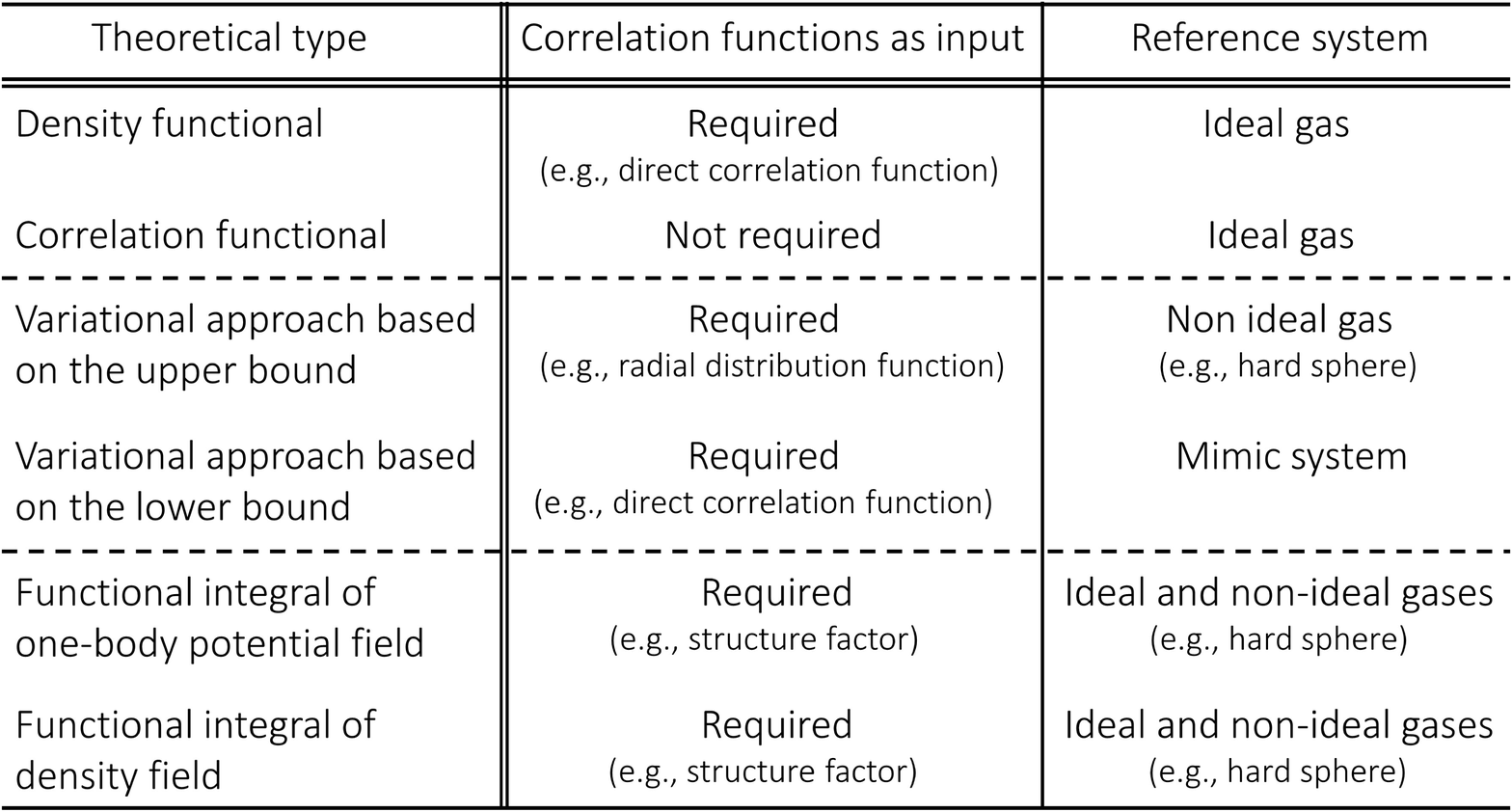}
\end{center}
\end{table*}

We can find a variety of relationships between the LST and the field theories for describing the liquid state.
To compare them, Table 1 classifies the previous field theories into three groups:
functional theory, variational theory (VT), and functional-integral (FI) representation.
First, the functional theory investigates the free energy functionals defined by the Legendre transforms, and is further divided into the density functional theory (DFT)  [11-14] and the correlation functional theory (CFT) [15-20] depending on the type of the Legendre transforms [20].
Second, the VT is based on the Gibbs-Bogoliubov inequalities [1, 2, 21, 22] and aim to optimize the free energy functional via tuning the two-body interaction potential of a reference system.
The difference in the reference system used leads to different bounds:
we have the VT based on the upper bound [21-31], and that on the lower bound [1, 32, 33, 80].
Last, the FI representations of the free energy functional integrate over the fields such as the one-body potential [34-43] and density fields [42, 44-56]. 

As seen from the second column of Table 1, it is only the CFT that produces by itself the above correlation functions (or the OZ equation and the approximate closure relation) without requiring the input from the LST [15-20].
Correspondingly, the third column of Table 1 lists what kind of reference systems have been set in developing the field theories.
When we consider a reference system of non-ideal gas, the bare interaction potential $v(r)$ needs to be divided into two parts:
\begin{eqnarray}
\beta v(r)=u(r)+w(r),
\label{separation}
\end{eqnarray}
where $\beta$ denotes the inverse of the thermal energy $k_BT$, and $u(r)$ corresponds to the interaction potential in the $k_BT$-unit of a reference fluid that we adopt.
Hard sphere fluids are one of the reference systems commonly used [1, 2, 24-28], and another well-known prescription for such potential separation has been provided by the systematic Weeks-Chandler-Andersen (WCA) perturbation theory for uniform fluids [1, 2, 29-31],
which has been successful for the interaction potentials with minima such as the Lennard-Jones potential.

Turning our attention to interfacial phenomena, including adsorption and wetting in nonuniform systems, the DFT incorporates attractive interactions, extracted from the above potential separation, in the mean-field approximation that neglects short-range correlations [11-14, 57-59]. 
Modern integral equation theories for inhomogeneous fluids provide an alternative approach to the DFT;
however, they usually suffer from problems of thermodynamic consistency [1-5].
Another promising approach is local molecular field theory developed by Weeks and co-workers [59-62].
Extending the idea of the above WCA theory for uniform fluids not only to nonuniform systems but also to the separation of interaction potentials without minima, the local molecular field theory has demonstrated that modified mean-field equation (MMF equation) is relevant to the description of various nonuniform fluids, including density distribution of counterions (inhomogeneous Coulomb fluids of point charges) in the strong coupling regime and electrostatics in models of confined water [59-62].

In terms of the potential separation, the local molecular field theory has the following implication.
There has been a problem that all species of fluids cannot set the reference system based on the conventional treatments mentioned above:
it is hard to find a reference system for pure hard sphere fluids themselves, and also the WCA separation is unavailable for the two-body potential with neither minimum nor a characteristic length of the potential profile such as the pure Coulomb interaction.
Nevertheless, the local molecular field theory suggests that we can successfully divide various bare interaction potentials $\beta v(r)$, including hard core potential and pure Coulomb potential, into strong short-ranged reference parts $u(r)$, and slowly varying long-ranged perturbation contributions $w(r)$.

In the MMF equation of the local molecular field theory, we use the long-ranged part $w(r)$, instead of the original one $\beta v(r)$, after choosing an optimal separation where the short-ranged potential $u(r)$ reproduces short-range correlations between the original fluid particles interacting via $\beta v(r)$ [57-62].
With the use of $w(r)$, the MMF equation has been found to precisely describe the inhomogeneous density distribution $\rho({\bf x})$ at the position ${\bf x}$ as follows [57-62]:
\begin{eqnarray}
&&\ln\left\{\frac{\rho({\bf x})}{\rho_B}\right\}=-\beta\phi_R({\bf x})\nonumber\\
&&\beta\phi_R({\bf x})=\beta\phi({\bf x})+\int d{\bf y}\,\left\{\rho({\bf y})-\rho_B\right\} w(|{\bf x}-{\bf y}|),\nonumber\\
\label{LMF}
\end{eqnarray}
where $\rho_B$ denotes the bulk density, and $\phi({\bf x})$ the external field which, for example, arises from a fixed solute.

The success of the MMF equation (\ref{LMF}) indicates that there is a criterion for separating every interaction potential into two parts without setting a priori any reference fluid;
at present, however, the function form of $u(r)$ (or $w(r)$) has been chosen on an ad hoc basis, and the adjustable parameter for the potential separation has been tuned to create a mimic system of particles interacting via $u(r)$ that can reproduce the radial distribution function $g(r)$ of the true system. 
In addition, it has been shown that the above mean-field DFT formally derives eq. (\ref{LMF}) using a given separation of the interactions potential, and yet loses the information on short-range correlations in mimic system of particles interacting via the short-ranged potential $u(r)$ [57-59]. 

Thus, this paper purposes both to validate such a modification of the mean-field equation field theoretically while considering short-range correlations, and to develop a systematic method of the potential separation that is available for any fluids, including hard spheres and Coulomb fluids of point charges.
We focus on the extension of the CFT to the FI form so that the field theoretic perturbation theory in the saddle-point (SP) approximation may apply to the description of fluctuating correlation fields in nonuniform liquids.

Such a field theoretic framework is increasingly becoming significant especially for glass-forming liquids where out-of-equilibrium correlations are to be investigated [63-79].
For example, the freezing of density fluctuations at the dynamical glass transition has been successfully reproduced by the replicated HNC approximation formulated in terms of a correlation function between two replicas which is related to an order parameter as an indicator of this transition [67-70].
Also, there are several issues to be addressed:
On one hand, it remains controversial as to whether or not density-density correlations in hard-particle packings toward jamming follow the behavior expected for random hyperuniform materials having an anomalous suppression of long-range density fluctuations [71-74].
On the other hand, ongoing development of simulation techniques raises the possibility that radial distribution functions $g(r)$ representing short-range correlations on a molecular scale are distinct for fluid and jammed states at the same density [75].
A combined approach of simultaneously considering short-and long-range correlations is therefore needed.

Our field theoretic formulation is to incorporate the FI of two-body fields (density-density correlation field and its dual interaction potential field) into the density FI (D-FI) representation, and determine the appropriate interaction potential of the long-ranged part $w(r)$, instead of $u(r)$, in the opposite direction of the local molecular field theory.
We evaluate the FI representation of density and two-body fields (the DT-FI form) in the SP approximation, thereby providing a set of self-consistent equations, equivalent to the MMF equation (\ref{LMF}). 

The remainder of this paper is organized as follows.
In section II, we formulate the Helmholtz free energy of the canonical system using both the VT based on the lower bound [1, 32, 33. 80] (the fourth row in Table 1) and the FI form of the one-body potential field based on the Hubbard-Stratonovich (HS) transformation [34-43] (the fifth row in Table 1) for the purpose of describing some problems that are faced in separating the interaction potential. 
In section III, it is demonstrated that an extra term remaining in the D-FI form creates inconsistency with the mean-field approximation based on the HS transformation, particularly in the canonical system.
In section IV, we address the fundamental issue of the D-FI form by introducing the FIs of two-body fields (both density-density correlation field and its dual field playing the role of two-body interaction potential).
Section V validates the DT-FI representation via investigating a set of self-consistent field equations with respect to the above two-body fields. 
It is to be noted that the SP method of the DT-FI is valid for any system, including hard sphere fluid and point charge system with purely Coulombic interactions (or the one component plasma with no hard core interactions). 
In section VI, the DT-FI theory in the mean-field approximation is compared both with the VT based on the lower bound [1, 32, 33, 80] that optimizes a trial free energy functional by its maximization, and with the VT based on the upper bound (or the WCA perturbation theory) [21-31].

\section{Three requirements for validating the modified mean-field (MMF) equation field theoretically}

There are some problems that arise when we separate the interaction potential without setting a priori any reference system.
To see this, we first write down the Helmholtz free energy of the canonical system that consists of $N$-particles interacting via the two-body interaction potential $v(|{\bf x}_i-{\bf x}_j|)$ between the $i$-th and the $j$-th particles located at ${\bf x}_i$ and ${\bf x}_j$, respectively.
It is convenient to introduce the instantaneous density-density correlation function $G_{\hat{\rho}}$ such that
\begin{eqnarray}
   G_{\hat{\rho}}({\bf x},{\bf y})=\hat{\rho}({\bf x})\hat{\rho}({\bf y})
   -\hat{\rho}({\bf x})\delta({\bf x}-{\bf y}),
\label{bare_correlation}
   \end{eqnarray}
where
\begin{equation}
\hat{\rho}({\bf x})=\sum_{i=1}^N\delta({\bf x}-{\bf x}_i).
\label{bare_density}
\end{equation}
The instantaneous density-density correlation function $G_{\hat{\rho}}$ allows to express the interaction energy $U\{\hat{\rho};\beta v\}$ of the simple form as follows:
\begin{eqnarray}
\beta U\{\hat{\rho};\beta v\}&=&\frac{1}{2}\int\int  d{\bf x}d{\bf y}
\,G_{\hat{\rho}}({\bf x},{\bf y})\beta v(|{\bf x}-{\bf y}|)\nonumber\\
&&\qquad\qquad\qquad\quad+\int d{\bf x}\,\beta\phi({\bf x})\hat{\rho}({\bf x}).
\label{u-start}
\end{eqnarray}
The Helmholtz free energy $F$ in the $k_BT$-unit is given by the configurational integral of $N$-particle positions, $\{{\bf x}_1,\cdots,{\bf x}_N\}$, as
\begin{eqnarray}
  e^{-F\{\beta v\}}&=&\mathrm{Tr}\,\exp\left[-\beta U\{\hat{\rho};\beta v\}\right]\nonumber\\
  \mathrm{Tr}&\equiv&
  \frac{\Lambda^{-3N}}{N!}\,\int d{\bf x}_1\cdots\int d{\bf x}_N
\label{f-start}
\end{eqnarray}
with the de Broglie wavelength $\Lambda$.

The VT based on the lower bound [1, 32, 33, 80] uses a variational functional $L\{w;g\}$ that depends on an arbitrary two-body interaction potential $w$ to be optimized relying on the lower bound of the true Helmholtz free energy $F\{\beta v\}$, instead of the usual upper bound.
In the potential separation such that eq. (\ref{separation}), we have a Helmholtz free energy $F\{w\}$ of a soft-core system interacting via unknown interaction potential $w$.
The Gibbs-Bogoliubov inequality states that the addition of a residual interaction energy provides the lower bound of the true free energy $F\{\beta v\}$ for a uniform liquid with the volume of $V$ and the smeared density of $\overline{\rho}\equiv N/V$ [1, 32, 33, 80]:
 \begin{eqnarray}
\frac{L\{w;g\}}{N}\equiv
\frac{F\{w\}}{N}+\frac{\overline{\rho}}{2}\int dr g(r)u(r)\leq
\frac{F\{\beta v\}}{N},
   \label{start} 
   \end{eqnarray}
where $\phi=0$ in eq. (\ref{u-start}) and $g(r)$ represents the exact radial distribution function of the true system.
Maximization of $L\{w;g\}$ with respect to $w$ yields
   \begin{eqnarray}
&&\left.
   \frac{1}{N}\left(\frac{\delta F\{w\}}{\delta w}\right)
   \right|_{w=w^*}=\frac{\overline{\rho}}{2}\,g(r)
\nonumber\\
&&\frac{L_{\mathrm{max}}\{w^*;g\}}{N}=
   \frac{\Gamma_2\{g\}}{N}+\frac{\overline{\rho}}{2}\int dr g(r)\beta v(r),
\label{def_L}
   \end{eqnarray}
where the correlation functional $\Gamma_2\{g\}$ is the second Legendre transform of $F\{w\}$:
   \begin{eqnarray}
\frac{\Gamma_2\{g\}}{N}&=&\frac{F\{w^*\}}{N}-\frac{\overline{\rho}}{2}\int dr g(r)w^*(r).
\label{def_gamma}
\end{eqnarray}
In the SP approximation of $F\{w\}$, maximization of $L\{w;g\}$ with respect to $w$ yields the OZ-like equation [32, 80]:
\begin{equation}
h(r)=-w^*(r)-\overline{\rho}\int d{\bf r}' 
w^*(|{\bf r}-{\bf r}'|)h(r'),
   \label{oz}
   \end{equation}
implying that
\begin{equation}
w^*(r)=-c(r).
\label{u_direct}
\end{equation}
It is to be noted that the inequality (\ref{start}) requires the exact form of $h(r)$ (or $g(r)$).
Conversely, any approximate form of $c(r)$ may be used in the VT based on the lower bound even though the equality (\ref{u_direct}) is verified within the SP approximation of $F\{w\}$.
In other words, the VT based on the lower bound lacks the capability to specify the approximation of the closure relation [1, 32, 80].

Next, we would like to see the conventional field theory based on the Hubbard-Stratonovich (HS) transformation [34-43].
As shown in the literature [34-43] (see also Appendix A), the SP equation of the HS form yields a mean-field equation as follows:
 \begin{eqnarray}
&&\ln\left\{\frac{\rho_{\mathrm{mf}}({\bf x})}{\rho_B}\right\}=-\psi_{\mathrm{mf}}({\bf x})\nonumber\\
&&\psi_{\mathrm{mf}}({\bf x})=\beta\phi({\bf x})+\int d{\bf y}\rho_{\mathrm{mf}}({\bf y})
\beta v(|{\bf x}-{\bf y}|)+\alpha^*\nonumber\\
&&\rho_{\mathrm{mf}}({\bf x})=\frac{N e^{-\psi_{\mathrm{mf}}({\bf x})}}{\int d{\bf x}\,e^{-\psi_{\mathrm{mf}}({\bf x})}},
\label{app_mf}
  \end{eqnarray}
where $\phi({\bf x})$ denotes the actual external field, the constant term $\alpha^*$ arises from the conservation of particle number, and $\rho_B=N/\int d{\bf x}\,e^{-\psi_{\mathrm{mf}}({\bf x})}$ the bulk density as before.
Equation (\ref{app_mf}) has been known to be the Poisson-Boltzmann equation when $v(|{\bf x}-{\bf y}|)$ is set to be the pure Coulomb potential $v(|{\bf x}-{\bf y}|)\sim 1/|{\bf x}-{\bf y}|$.
For the singular hard core potential, on the other hand, the HS representation is invalid and the associated mean-field equation (\ref{app_mf}) is irrelevant to find the density distribution $\rho_{\mathrm{mf}}$.
Hence, the field theories have adopted perturbation forms around a non-ideal gas such as hard sphere fluid for taking into account the short-range correlations due to the hard-core interaction as input, which is traced back to the Hubbard-Schofield transformation [36-41] (not the Hubbard-Stratonovich one).

Our focus, however, is on the development of the field theoretic method that replaces the total of the bare interaction potential $v$ in eq. (\ref{app_mf}) by an appropriate interaction potential as given by eq. (\ref{LMF}). 
In Appendix A, we show that incorporation of two-body fields into the HS form yields a mean-field equation modified as
  \begin{eqnarray}
&&\ln\left\{\frac{\rho'_{\mathrm{mf}}({\bf x})}{\rho_B}\right\}=-\psi'_{\mathrm{mf}}({\bf x})\nonumber\\
&&\psi'_{\mathrm{mf}}({\bf x})=\beta\phi({\bf x})-\int d{\bf y}\,\rho'_{\mathrm{mf}}({\bf y})
W^*(|{\bf x}-{\bf y}|)+\alpha^*\nonumber\\
&&\rho'_{\mathrm{mf}}({\bf x})=\frac{N e^{-\psi'_{\mathrm{mf}}({\bf x})}}{\int d{\bf x}\,e^{-\psi'_{\mathrm{mf}}({\bf x})}},
\label{app_mf2}
  \end{eqnarray}
where $-W^*$ corresponds to $w$ in eq. (\ref{LMF});
however, the interaction potential $-W^*$ cannot be fixed within the SP approximation of two-body fields (see Appendix A).

We have seen so far that both treatments of the VT based on the lower bound and the HS form with two-body fields introduced are unable to provide the closure relations for specifying the approximate form of the optimized interaction potential ($w^*$ or $-W^*$):
while the radial distribution function $g(r)$ in the VT is not a variable function but is fixed at that of the true system and the associated direct correlation function $c(r)$ should in principle be exact, the SP approximations of two-body fields in the HS representation give inconsistent results of $-W^*$ (see Appendix A for the details).
In order to validate the MMF equation, such as eq. (\ref{app_mf2}) with the interaction potential $-W^*$ determined, we need to develop a new field theory that satisfies the following requirements:
\begin{description}
\item[(R1)] Whether a potential separation is performed or not, the obtained mean-field equation has the same form as eqs. (\ref{LMF}), (\ref{app_mf}) and (\ref{app_mf2}).
\item[(R2)] In the SP approximation of the Helmholtz free energy, the optimized interaction potential ($w^*(r)$ or $-W^*(r)$) in the $k_BT$-unit is given by minus the direct correlation function, similarly to the OZ-like equation (\ref{oz}).
\item[(R3)] The DT-FI representation itself provides an approximate closure relation in addition to the OZ-like equation, as well as the CFT. 
\end{description}
This paper demonstrates that these requirements are met by starting with the D-FI representation [42, 44-56], instead of the HS transformation.

\section{Density functional integral (D-FI) representation}
\subsection{The SP approximation of one-body potential field $\psi$}

As found from Appendix A, the FI representation of eq. (\ref{f-start}) is given by
\begin{eqnarray}
&&e^{-F\{\beta v\}}=\int D\psi\int D\rho\,\Delta_N\nonumber\\
&&\qquad\qquad\qquad\times\mathrm{Tr}\,
e^{\int d{\bf x}i\psi({\bf x})
  \left\{\hat{\rho}({\bf x})-\rho({\bf x})\right\}-\beta U\{\rho;\beta v\}}\nonumber\\
&&=\int D\psi\int D\rho\,\Delta_N
\>e^{-\beta\left[U\{\rho;\beta v\}-TS\{\rho;\psi\}\right]},
\label{FI-rho-psi}
\end{eqnarray}
where $\Delta_N=\delta\left[\int d{\bf x}\,\rho({\bf x})-N\right]$, the interaction energy $U\{\rho;\beta v\}$ now depends on the $\rho$-field and $S\{\rho;\psi\}$ denotes the entropic contribution expressed as
\begin{eqnarray}
&&-\frac{S}{k_B}\{\rho;\psi\}=
\int d{\bf x}i\psi({\bf x})\rho({\bf x})-\ln\left\{\mathrm{Tr}\,
e^{\int d{\bf x}i\psi({\bf x})\hat{\rho}({\bf x})}\right\}
\nonumber\\
&&=\int d{\bf x}i\psi({\bf x})\rho({\bf x})-N\ln\left\{\frac{\int d{\bf x}e^{i\psi({\bf x})}}{N}\right\}-N.
\nonumber\\
\label{bare-s}
\end{eqnarray}
Here we obtain the D-FI representation from integrating out the one-body potential field $\psi({\bf x})$ in the SP approximation [44, 46-51], though the D-FI theory going beyond the SP approximation of the $\psi$-field has been developed [42, 52-56].
There are two steps: 
(i) we obtain the solution of the SP equation, and (ii) we evaluate the fluctuations around the SP field in the Gaussian approximation, the so-called SP approximation.

First, the SP equation yields the relation, $\delta S/\delta\psi|_{\psi=i\psi^*}=0$.
The SP field $i\psi^*$ must be purely imaginary [42-55], so that we have
\begin{equation}
\rho({\bf x})=\frac{N e^{-\psi^*({\bf x})}}{\int d{\bf x}\,e^{-\psi^*({\bf x})}}.
\label{sp-equation}
\end{equation}
Expanding the entropic contribution $S$ around the SP field up to the quadratic term, the entropic contribution given by eq. (\ref{bare-s}) is written as
\begin{eqnarray}
&&S\{\rho;\psi\}\approx
S\{\rho;i\psi^*\}+\int d{\bf x}
\left.\frac{\delta S}{\delta\psi}\right|_{\psi=i\psi^*}\Delta\psi({\bf x})
\nonumber\\
&&
\left.+\frac{1}{2}\int\int d{\bf x}d{\bf y}\frac{\delta^2 S}{\delta\psi({\bf x})\delta\psi({\bf y})}
\right|_{\psi=i\psi^*}
\Delta\psi({\bf x})\Delta\psi({\bf y}),
\nonumber\\
\label{s-expansion1}
\end{eqnarray}
where $-S/k_B$ along the SP path gives the ideal entropy term $S_{\mathrm{id}}$:
\begin{eqnarray}
-\frac{S_{\mathrm{id}}\{\rho\}}{k_B}&=&-\frac{S}{k_B}\{\rho;i\psi^*\}\nonumber\\
&=&\int d{\bf x}\left\{\rho({\bf x})\ln\rho({\bf x})-\rho({\bf x})\right\},
\label{s-mean}
\end{eqnarray}
the second term on the right hand side (rhs) of eq. (\ref{s-expansion1}) vanishes due to the SP equation (\ref{sp-equation}), and $\Delta\psi({\bf x})$ in the last term on the rhs of eq. (\ref{s-expansion1}) is now the complex fluctuation field: $\Delta\psi=\Delta\psi_R+i\Delta\psi_I$.
In eq. (\ref{s-expansion1}), the second derivative of $S$ arises from the second term in the second line of eq. (\ref{bare-s}), providing
   \begin{eqnarray}
&&\left.\frac{\delta^2 (-S/k_B)}{\delta\psi({\bf x})\delta\psi({\bf y})}\right|_{\psi=i\psi^*}\nonumber\\
&&=\frac{N e^{-\psi^*({\bf x})}}{\int d{\bf x}\,e^{-\psi^*({\bf x})}}\delta({\bf x}-{\bf y})
-\frac{N e^{-\psi^*({\bf x})}}{\int d{\bf x}\,e^{-\psi^*({\bf x})}}\left\{
\frac{e^{-\psi^*({\bf y})}}{\int d{\bf y}\,e^{-\psi^*({\bf y})}}
\right\}\nonumber\\
&&=\gamma\rho({\bf x})\delta({\bf x}-{\bf y})-\frac{G_{\rho}({\bf x},{\bf y})}{N},
\label{s-second}
  \end{eqnarray}
where $\gamma=1-1/N$, and the second terms in the second and third lines of eq. (\ref{s-second}) correspond to correction terms due to the canonical system and is not negligible especially for ${\bf x}\neq{\bf y}$.
Combining eqs. (\ref{s-expansion1}), (\ref{s-mean}) and (\ref{s-second}), we obtain
\begin{eqnarray}
-\frac{S}{k_B}\{\rho;\psi\}&\approx&-\frac{S_{\mathrm{id}}\{\rho\}}{k_B}+\frac{1}{2}\int\int d{\bf x}
\,\gamma\rho(\bf x)\Delta\psi({\bf x})^2\nonumber\\
&&-\frac{1}{2N}\int\int d{\bf x}d{\bf y}G_{\rho}({\bf x},{\bf y})
\Delta\psi({\bf x})\Delta\psi({\bf y}).
\nonumber\\
\label{s-expansion2}
\end{eqnarray}
The last two terms on the rhs of eq. ({\ref{s-expansion2}) indicate that the SP configurations are different between the cases of ${\bf x}={\bf y}$ and ${\bf x}\neq{\bf y}$:
For ${\bf x}={\bf y}$, the local minimum and maximum are located at the SP filed along the real and imaginary fields ($\Delta\psi_R$ and $\Delta\psi_I$), respectively, and vice versa for ${\bf x}\neq{\bf y}$.

\subsection{On the requirement (R1)}

Based on the requirement (R1) described at the end of section II, we compare the SP path of the $\rho$-field in the D-FI form of eq. (\ref{FI-rho-psi}) with eq. (\ref{app_mf}) obtained from the HS form.
It is verified below that eq. (\ref{app_mf}) is equivalent to the following relation [45-50]:
\begin{equation}
\left.
\frac{\delta}{\delta\rho}\left\{
\beta(U-TS_{id})-\lambda\left(\int d{\bf x}\,\rho({\bf x})-N
\right)
\right\}\right|_{\rho=\rho_{\mathrm{mf}}}=0,
\label{general-mf}
\end{equation}
where the constraint $\Delta_N$ is considered using the Lagrange multiplier $\lambda$.
Since $\beta(U-TS_{id})$ in eq. (\ref{general-mf}) is written as
\begin{eqnarray}
&&\beta(U-TS_{id})=\frac{1}{2}\int\int d{\bf x}d{\bf y}\,\rho({\bf x})\rho({\bf y})\beta v(|({\bf x}-{\bf y}|)\nonumber\\
&&+\int d{\bf x}\,\left\{\beta\phi({\bf x})\rho({\bf x})+\rho({\bf x})\ln\rho({\bf x})\right\}-N-\frac{N}{2}\beta v(0),\nonumber\\
\label{general_mf2}
\end{eqnarray}
we have
\begin{eqnarray}
\ln\rho_{\mathrm{mf}}({\bf x})=-\beta\phi({\bf x})-\int d{\bf y}\,\beta v(|{\bf x}-{\bf y}|)\rho_{\mathrm{mf}}({\bf y})
+\lambda-1.\nonumber\\
\label{rho_mf}
\end{eqnarray}
Furthermore, comparison between eqs. (\ref{LMF}) and (\ref{rho_mf}) indicates that eq. (\ref{rho_mf}) becomes similar to eq. (\ref{LMF}) when
\begin{eqnarray}
\lambda-1&=&\ln\rho_B+\int d{\bf y}\,\beta v(|{\bf x}-{\bf y}|)\rho_B\nonumber\\
\rho_B&=&\frac{N}{\int d{\bf x}\,e^{-\beta\phi_R({\bf x})}}\nonumber\\
\beta\phi_R({\bf x})&=&\beta\phi({\bf x})+\int d{\bf y}\,\left\{\rho_{\mathrm{mf}}({\bf y})-\rho_B\right\}
\beta v(|{\bf x}-{\bf y}|),\nonumber\\
\label{lambda}
\end{eqnarray} 
where we have introduced the symbol $\beta\phi_R$ in order to represent the effective external field instead of $\psi_{\mathrm{mf}}$, other than eq. (\ref{app_mf}), so that it is clarified that the effective external field depends not on $\psi$ as a field variable in the HS form, but on $\rho$.
Combining eqs. (\ref{rho_mf}) and (\ref{lambda}), we have
\begin{eqnarray}
&&\ln\left\{\frac{\rho_{\mathrm{mf}}({\bf x})}{\rho_B}\right\}=-\beta\phi_R({\bf x})\nonumber\\
&&\beta\phi_R({\bf x})=\beta\phi({\bf x})+\int d{\bf y}\,\left\{\rho_{\mathrm{mf}}({\bf y})-\rho_B\right\} \beta v(|{\bf x}-{\bf y}|),\nonumber\\
\label{rho_mf2}
\end{eqnarray}
similarly to the mean-field equation (\ref{app_mf}) in the HS form and to the LMF equation (\ref{LMF}), though $\beta v$ needs to be replaced by $w$ in eq. (\ref{LMF}).
It is also noted that the Lagrange multiplier given by eq. (\ref{lambda}) actually assures the constraint $\Delta_N$:
\begin{eqnarray}
&&\int d{\bf x}\,\rho_{\mathrm{mf}}({\bf x})=N\nonumber\\
&&\rho_{\mathrm{mf}}({\bf x})=\rho_Be^{-\beta\phi_R({\bf x})}=\frac{Ne^{-\beta\phi_R({\bf x})}}{\int d{\bf x}\,e^{-\beta\phi_R({\bf x})}},
\label{rho-constraint}
\end{eqnarray}
as found from eq. (\ref{rho_mf2}).

The above discussions reveal that the present form of the D-FI does not meet the first requirement (R1), contradicting the conventional field theory:
discrepancy between the HS form and the D-FI representation necessarily exists unless the last two terms on the rhs of eq. (\ref{s-expansion2}) somehow disappear.
In what follows, we demonstrate that a solution to this problem of (R1) results in satisfying the other requirements (R2) and (R3). 

\section{Two-body fields incorporated into the D-FI representation}
\subsection{Constraints on the instantaneous density-density correlation field}

Before addressing the above extra terms in the entropic contribution of eq. (\ref{s-expansion2}), let us incorporate the instantaneous density-density correlation field $G=G_{\rho}$ into the D-FI form.
What is lacking in introducing the density-density correlation field $G$ (see Appendix A in eq. (\ref{app_g-identity})) is to ensure the positivity that $G\geq 0$;
while the positivity, $\rho\geq 0$, of the density field in the D-FI representation is satisfied as a result of the existence of the logarithmic term in $S_{\mathrm{id}}$, the FI range of both the $\rho$- and $G$-fields has no restriction and it is indispensable to fix the sign of $G$ in a natural way so that we can develop a DT-FI representation that is physically relevant.
To resolve the sign problem, we relate the correlation field $G$ in eqs. (\ref{app_g-identity}) and (\ref{app_g-fourier}) to the auxiliary field $\mathcal{M}$ for ${\bf x}\neq{\bf y}$:
  \begin{equation}
   G({\bf x},{\bf y})=\mathcal{M}^2({\bf x},{\bf y})\geq 0\quad({\bf x}\neq{\bf y}).
\label{positivity}
  \end{equation}
Furthermore, the inherent definition of the bare correlation function $G_{\hat{\rho}}$ given by eq. (\ref{bare_correlation}) imposes the restriction that $G({\bf x},{\bf y})=0$ at ${\bf x}={\bf y}$.

To take into account these constraints, we introduce the identity
  \begin{eqnarray}
&&1=\int DG\int D\mathcal{M}\,\left|\,\mathrm{det}\mathcal{M}\right|\nonumber\\
&&\qquad\qquad\times
\prod_{{\bf x}(\neq{\bf y})} \prod_{{\bf y}}\>
\delta\left[
  \frac{\mathcal{M}^2({\bf x},{\bf y})}{2}
-\frac{G({\bf x},{\bf y})}{2}
   \right]
\nonumber\\
&&\qquad\qquad
\times
\prod_{{\bf x}} \delta\left[G({\bf x},{\bf x})\right]\>
\prod_{{\bf x}}\prod_{{\bf y}} \delta\left[G({\bf x},{\bf y})-G_{\rho}({\bf x},{\bf y})\right]
\nonumber\\
&&=\int DG\int DW\int D\mathcal{M}\,\,\left|\,\mathrm{det}\mathcal{M}\right|\nonumber\\
&&\quad\>\times
\prod_{{\bf x}(\neq{\bf y})} \prod_{{\bf y}}\>
\delta\left[
  \frac{\mathcal{M}^2({\bf x},{\bf y})}{2}
-\frac{G({\bf x},{\bf y})}{2}
   \right]\delta[G({\bf x},{\bf x})]\nonumber\\
&&\quad\>\times\exp\left[\int \int d{\bf x}d{\bf y}\frac{iW({\bf x},{\bf y})}{2}
   \left\{G({\bf x},{\bf y})-G_{\rho}({\bf x},{\bf y})\right\}
   \right],\nonumber\\
\label{m-identity}
   \end{eqnarray}
through which it is satisfied that $G({\bf x},{\bf y})\geq 0\>({\bf x}\neq{\bf y})$ and $G({\bf x},{\bf x})=0$.
In eq. (\ref{m-identity}), we have used that the $b$-field as a function of $a$ such that $f(a)=b$ has the following relation:
 \begin{eqnarray}
   1&=&\int\,Db\,\prod_{{\bf x}}\>\delta\left[
   b({\bf x})-\hat{b}({\bf x})
   \right]\nonumber\\
  &=&\int\,Da\,\left|\,\mathrm{det}
   \frac{\delta f(a)}{\delta a}\right|
   \prod_{{\bf x}}\>\delta\left[
   f[a({\bf x})]-\hat{b}({\bf x})
   \right].
\label{det-identity}
   \end{eqnarray}

Equation (\ref{FI-rho-psi}) multiplied by the rhs of eq. (\ref{m-identity}) leads to the following form:
\begin{eqnarray}
 &&e^{-F\{\beta v\}}=\int DW\int D\mathcal{M}\int D\psi\int D\rho\,\Delta_N
\left|\,\mathrm{det}\mathcal{M}\right|
e^{-\mathcal{L}}
\nonumber\\
&&\mathcal{L}\{\rho;\psi\}=\beta\left[U\{\rho;iW\}-TS\{\rho;\psi\}\right]\nonumber\\
&&\qquad+\frac{1}{2}\int\int_{{\bf x}\neq{\bf y}} d{\bf x} d{\bf y}\mathcal{M}^2({\bf x},{\bf y})\left\{
\beta v(|{\bf x}-{\bf y}|)-iW({\bf x},{\bf y})\right\}.
\nonumber\\
\label{psi-start}
\end{eqnarray}
Going back to eq. (\ref{s-expansion2}), we would like to note that $G_\rho$ can be replaced by $\mathcal{M}^2$ thanks to the delta functionals of eq. (\ref{m-identity}).
It follows that the potential integral contribution is written as
\begin{eqnarray}
&&\int D\psi\,e^{S/k_B\{\rho;\psi\}}\approx
e^{S_{\mathrm{id}}/k_B\{\rho\}}\nonumber\\
&&\qquad\qquad\qquad\times\int D(\Delta\psi_R)e^{\Delta S_R/k_B}\int D(\Delta\psi_I)e^{\Delta S_I/k_B}\nonumber\\
&&\frac{\Delta S_R}{k_B}=-\frac{1}{2}\int d{\bf x}\,\rho({\bf x})\Delta\psi_R^2({\bf x})\nonumber\\
&&\frac{\Delta S_I}{k_B}=-\frac{1}{2N}\int\int_{{\bf x}\neq{\bf y}} d{\bf x}d{\bf y}\mathcal{M}^2({\bf x},{\bf y})
\Delta\psi_I({\bf x})\Delta\psi_I({\bf y}),\nonumber\\
\label{s-expansion3}
\end{eqnarray}
where not the approximation but the constraint $G({\bf x},{\bf x})=0$ imposed by the delta functional in eq. (\ref{m-identity}) is considered.
The Gaussian integrations yield
\begin{eqnarray}
\int D\psi\,e^{S/k_B\{\rho;\psi\}}=e^{S_{\mathrm{id}}/k_B\{\rho\}}
\mathrm{det}\sqrt{\frac{2\pi}{\rho}}\,\mathrm{det}\frac{\sqrt{2N\pi}}{|\mathcal{M}|},\nonumber\\
\label{psi-det}
\end{eqnarray}
which reads
\begin{eqnarray}
|\mathrm{det}\mathcal{M}|\int D\psi\,e^{S/k_B\{\rho;\psi\}}
=e^{S_{\mathrm{id}}/k_B\{\rho\}}e^{-\frac{1}{2}\ln\mathrm{det}\rho},
\end{eqnarray}
where the determinant term $|\mathrm{det}\mathcal{M}|$ has been canceled by the last factor, $\mathrm{det}(\sqrt{2N\pi}/|\mathcal{M}|)$, on the rhs of eq. (\ref{psi-det}) and constant terms have been omitted for simplicity.
As shown previously [46], the correction term $(1/2)\ln(\mathrm{det}\rho)$ is written as
\begin{eqnarray}
\frac{1}{2}\ln\mathrm{det}\rho&=&\int d{\bf x}\lim_{a\rightarrow 0}\frac{1}{2a^3}\ln(\rho_la^3)\nonumber\\
\lim_{a\rightarrow 0}\frac{1}{a^3}\ln(\rho_la^3)&=&\lim_{a\rightarrow 0}\left(\rho_l-\overline{\rho}+\mathcal{O}[a^3]\right)
=\rho({\bf x})-\overline{\rho},\nonumber\\
\label{log}
\end{eqnarray}
where $a$ denotes the lattice constant defined by $\overline{\rho}a^3=1$, we have used the discretized expression of the Gaussian integration while taking the continuum limit of $a\rightarrow 0$, and the logarithmic expansion have been performed as $\ln(\rho_la^3)=\ln\{1+(\rho_l-\overline{\rho})a^3\}\approx(\rho_l-\overline{\rho})a^3+\mathcal{O}[\{(\rho_l-\overline{\rho})a^3\}^2]$.
It follows from eq. (\ref{log}) that the correction term, $(1/2)\ln(\mathrm{det}\rho)$, is negligible:
$(1/2)\ln(\mathrm{det}\rho)=0.5(N-N)=0$.

Thus, we obtain
\begin{eqnarray}
&&e^{-F\{\beta v\}}=\int DW\int D\mathcal{M}\int D\rho\,\Delta_N\,e^{-\mathcal{L}}
\nonumber\\
&&\mathcal{L}=\mathcal{A}_{\mathrm{mf}}\{\rho;iW\}\nonumber\\
&&\qquad+\frac{1}{2}\int\int_{{\bf x}\neq{\bf y}} d{\bf x} d{\bf y}\mathcal{M}^2({\bf x},{\bf y})\left\{
\beta v(|{\bf x}-{\bf y}|)-iW({\bf x},{\bf y})\right\}.
\nonumber\\
&&\mathcal{A}_{\mathrm{mf}}\{\rho;iW\}=\beta\left\{
U\{\rho;iW\}-TS_{\mathrm{id}}\{\rho\}
\right\},
\label{psi-result}
\end{eqnarray}
where the density functional $\mathcal{A}_{\mathrm{mf}}\{\rho;iW\}$ is of the form of the Helmholtz free energy type in the mean-field approximation.
Remarkably, the extra terms arising from the entropic contribution (see eq. (\ref{s-expansion3})) vanishes in eq. (\ref{psi-result}) due to the introduction of auxiliary field $\mathcal{M}$.

A few remarks are in order about the cancellation of the determinant factor $|\mathrm{det}\mathcal{M}|$ due to both the SP approximation of the $\psi$-field and the introduction ot the auxiliary $\mathcal{M}$-field:
\begin{itemize}
\item The positivity of $G(=G_{\rho})$ can be imposed by other functions such that $e^{\mathcal{M}}=G$;
the present relation $\mathcal{M}^2=G$, however, has been found appropriate for the SP approximation of the $\psi$-field.
\item The determinant factor $|\mathrm{det}\mathcal{M}|$ in eq. (\ref{m-identity}) has been canceled due to the extra terms given by eq. (\ref{s-expansion3}) but would inevitably remains in the SP equation of the HS form even if the auxiliary field $\mathcal{M}$ is introduced for ensuring the positivity of $G$, which is the reason why we have adopted the D-FI representation instead of the HS form.
\item We have neglected higher order contributions to the DT-FI form (\ref{psi-result}) beyond the SP approximation of the $\psi$-field, which implies that the expression of the DT-FI varies in accordance with the approximations of the $\psi$-field;
we should try to improve the accuracy in terms of both the $\psi$-and $\rho$-fields when going beyond the SP approximation along the line of previous formulations [52-56].
\end{itemize}
These clarify that the D-FI form of the canonical system has prescribed the DT-FI representation, the hybrid field theory of the density and two-body fields, that allows to impose the constraints on density-density correlation field (i.e., $G({\bf x},{\bf y})\geq 0\>({\bf x}\neq{\bf y})$ and $G({\bf x},{\bf x})=0$).
In the first step, we will evaluate the $\rho$-field within the SP approximation of the DT-FI representation, according to the last remark made above.

\subsection{The Helmholtz free energy functional prior to optimization of two-body fields}

Equation ({\ref{psi-result}}) implies that the MMF equation (\ref{app_mf2}) can be obtained from the SP equation with respect to the $\rho$-field:
\begin{eqnarray}
\left.
\frac{\delta\mathcal{A}_{\mathrm{mf}}\{\rho;iW\}}{\delta\rho}\right|_{\rho=\rho^*}=0,
\label{sp-a-rho}
\end{eqnarray}
which actually reads
 \begin{eqnarray}
&&\ln\left\{\frac{\rho^*({\bf x})}{\rho_B}\right\}=-\beta\phi_R({\bf x})\nonumber\\
&&\beta\phi_R({\bf x})=\beta\phi({\bf x})+\int d{\bf y}\,\left\{\rho^*({\bf y})-\rho_B\right\}
iW(|{\bf x}-{\bf y}|)\nonumber\\
&&\rho^*({\bf y})=\frac{N e^{-\beta\phi_R({\bf y})}}{\int d{\bf y}\,e^{-\beta\phi_R({\bf y})}},
\label{sp-mmf}
  \end{eqnarray}
similarly to eq. (\ref{rho_mf2}).
In the SP approximation of the $\rho$-field, we consider the quadratic density fluctuation term around $\rho^*$:
\begin{eqnarray}
&&\mathcal{A}_{\mathrm{mf}}\{\rho;iW\}\approx
\mathcal{A}_{\mathrm{mf}}\{\rho^*;iW\}\nonumber\\
&&+\frac{1}{2}\int\int d{\bf x}d{\bf y}\,\left\{
iW({\bf x},{\bf y})+\frac{\delta({\bf x}-{\bf y})}{\rho^*({\bf x})}\right\}\Delta\rho({\bf x})\Delta\rho({\bf y}),
\nonumber\\
\label{rho-expansion}
\end{eqnarray}
where $\Delta\rho=\rho-\rho^*$.

The Gaussian integration over the $\Delta\rho$-field yields
\begin{eqnarray}
&&e^{-\beta F\{\beta v\}}=\int DW\int D\mathcal{M}\>e^{-\mathcal{L}_{\mathrm{RPA}}}\nonumber\\
&&\mathcal{L}_{\mathrm{RPA}}=\mathcal{A}\{\rho^*;iW\}
+\frac{1}{2}\ln\mathrm{det}\left\{
iW({\bf x},{\bf y})+\frac{\delta({\bf x}-{\bf y})}{\rho^*({\bf x})}
\right\}
\nonumber\\
&&\quad+\frac{1}{2}\int\int_{{\bf x}\neq{\bf y}} d{\bf x}d{\bf y}\,\mathcal{M}^2({\bf x},{\bf y})\left\{
\beta v(|{\bf x}-{\bf y}|)-iW({\bf x},{\bf y})\right\}.
\nonumber\\
\label{rpa}
\end{eqnarray}
While the functional of $\mathcal{L}_{\mathrm{RPA}}$ is similar to the conventional free energy functional in the random phase approximation (RPA) of the LST [1, 6-10, 24-28], both of the correlation field $\mathcal{M}^2$ and the two-body interaction potential field $iW$ are arbitrary in eq. (\ref{rpa}) and remains to be determined.

\subsection{Self-consistent equations of two-body fields and associated results}

It is necessary to evaluate the FIs of two fields, $\mathcal{M}$ and $W$, for obtaining the final form of the Helmholtz free energy from eq. (\ref{rpa}).
In this paper, we limit our discussion both to the SP paths of $\mathcal{M}$ and $W$ and to the use of the functional of the RPA type (the RPA functional) given by (\ref{rpa}):
\begin{eqnarray}
\left.\frac{\delta\mathcal{L}_{\mathrm{RPA}}}{\delta\mathcal{M}}\right|_{\mathcal{M}={\mathcal{M}}^*}&=&0
\label{sp-m}\\
\left.\frac{\delta\mathcal{L}_{\mathrm{RPA}}}{\delta W}\right|_{W=iW^*}&=&0,
\label{sp-w}
\end{eqnarray}
where the solution of $W$ has been set to be purely imaginary ($W=iW^*$) so that $iW=-W^*$ may provide real field as an optimized two-body potential physically meaningful.

Equation (\ref{sp-m}) reads
\begin{eqnarray}
\mathcal{M}^*({\bf x},{\bf y})\left\{
\beta v(|{\bf x}-{\bf y}|)-iW({\bf x},{\bf y})\right\}=0,
\label{m-meanfield}
\end{eqnarray}
which holds for any approximations to the DT-FI including the above RPA.
Equation (\ref{sp-w}), on the other hand, is rewritten by assuming the existence of the following function $k({\bf x},{\bf y})$:
\begin{equation}
{M}^*({\bf x},{\bf y})^2-\rho^*({\bf x})\rho^*({\bf y})=\rho^*({\bf x})\rho^*({\bf y})k({\bf x},{\bf y}),
\label{assumption}
\end{equation}
with use of which eq. (\ref{sp-w}) reads when using the approximate form (\ref{rpa}):
\begin{eqnarray}
k({\bf x},{\bf y})=W^*({\bf x},{\bf y})+\int d{\bf z}\,
k({\bf x},{\bf z})\rho^*({\bf z})W^*({\bf z},{\bf y}),
\label{oz2}
\end{eqnarray}
similarly to the OZ-like equation (\ref{oz}) in the case of nonuniform fluids (see Appendix B for the detailed derivation).
It follows that the two-body fields can be related to the correlation functions (the total correlation function $h^*(r)$, the radial distribution one $g^*(r)$, and the direct correlation one $c^*(r)$) such that
\begin{eqnarray}
k({\bf x},{\bf y})&=&h^*(|{\bf x}-{\bf y}|)\nonumber\\
{M}^*({\bf x},{\bf y})^2
&=&\rho^*({\bf x})\rho^*({\bf y})g^*(|{\bf x}-{\bf y}|)\nonumber\\
W^*({\bf x},{\bf y})&=&c^*(|{\bf x}-{\bf y}|),
\label{twobody-solution}
\end{eqnarray}
where the superscript * denotes that these correlation functions are defined solely by the relations (\ref{m-meanfield}) and (\ref{oz2}).
Equation (\ref{twobody-solution}) transforms eq. (\ref{m-meanfield}) to the relation,
\begin{equation}
\sqrt{g^*(r)}\left\{\beta v(r)+c^*(r)\right\}=0,
\label{smsa}
\end{equation}
which has the same solution as that of the MSA closure relation: $g^*(r)=0$ for $r\leq\sigma_{\mathrm{eff}}$ and $\beta v(r)+c^*(r)=0$ for $r>\sigma_{\mathrm{eff}}$ with $\sigma_{\mathrm{eff}}$ denoting the effective diameter, as well as the MSA [1-10].

To summarize, the RPA functional (\ref{rpa}) leads to a set of self-consistent equations (\ref{m-meanfield}) to (\ref{smsa}) for the two-body fields, thereby validating not only the MMF equation but also the RPA functional that applies to any systems including hard sphere fluids and Coulomb fluids with no hard core such as the one component plasma of point charges.
The MMF equation (\ref{sp-mmf}) is thus given by
\begin{eqnarray}
&&\ln\left\{\frac{\rho_{\mathrm{mf}}({\bf x})}{\rho_B}\right\}=-\beta\phi_R({\bf x})\nonumber\\
&&\beta\phi_R({\bf x})=\beta\phi({\bf x})-\int d{\bf y}\,\left\{\rho_{\mathrm{mf}}({\bf y})-\rho_B\right\} c^*(|{\bf x}-{\bf y}|),\nonumber\\
\label{rho_mmf3}
\end{eqnarray}
which is nothing but the LMF equation (\ref{LMF}) with setting that $-c^*=w$.

When we consider $\rho^*=\overline{\rho}$ in the absence of external field ($\phi=0$), the RPA functional (\ref{rpa}) combined with the SP solutions (eqs. (\ref{m-meanfield}) to (\ref{smsa})) of the two-body fields yields
 \begin{eqnarray}
   \frac{F\{\beta v\}}{V}
&=&\frac{\overline{\rho}}{2}c(0)-\frac{\overline{\rho}^2}{2}\int dr\,c^*(r)
+\overline{\rho}\ln\overline{\rho}-\overline{\rho}
\nonumber\\
&&\qquad\qquad+\frac{1}{2}\int_{{\bf k}}\ln\left[1-\overline{\rho}c^*({\bf k})\right],
\label{lower-result}
   \end{eqnarray}
which corresponds to an extended form of the optimized RPA functional [1, 2, 6-10, 24-28] that requires no reference fluid and accordingly adds no perturbative interaction potential [6-10].

It is also to be noted that the interaction energy term $U\{\rho^*;-c^*\}$ in the mean-field functional $\mathcal{A}_{\mathrm{mf}}\{\rho^*;-c^*\}$, which corresponds to the first two terms on the rhs of eq. (\ref{lower-result}), retains the information on the short-range correlations as found from the expression as follows:
 \begin{eqnarray}
&&\beta U\{\rho^*;-c^*\}\nonumber\\
&&=-\frac{1}{2}\int\int d{\bf x}d{\bf y}\,\rho^*({\bf x})\rho^*({\bf y})g^*(|{\bf x}-{\bf y}|)c^*(|{\bf x}-{\bf y}|)\nonumber\\
&&\quad-\frac{1}{2}\int d{\bf x}\,\rho^*({\bf x}),
\label{mf-energy}
   \end{eqnarray}
where use has made of the following relations due to the OZ-like equation (\ref{oz2}):
\begin{eqnarray}
h^*(0)&=&c^*(0)+\int d{\bf y}\,
g(|{\bf x}-{\bf y}|)\rho^*({\bf y})c^*(|{\bf y}-{\bf x}|)
\nonumber\\
&&\qquad-\int d{\bf y}\,
\rho^*({\bf y})c^*(|{\bf x}-{\bf y}|),\nonumber\\
h^*(0)&=&-1.
\label{oz3}
\end{eqnarray}
Equation (\ref{mf-energy}) reveals that the mean-field free energy functional $\mathcal{A}_{\mathrm{mf}}\{\rho^*;-c^*\}$ contains the information on the short-range correlations through $g^*(r)$, despite being the mean-field approximation.

\section{Comparison between the DT-FI form and the variational theory (VT) based on the lower and upper bounds}

There are a couple of features that is shared by the DT-FI form and the VT based on the lower bound [1, 32, 33, 80], which will be described first for uniform fluids in the absence of external field $\phi=0$.

In terms of the second Lengendre transform, it is seen that the RPA functional $\mathcal{L}_{\mathrm{RPA}}$ prior to the optimization, eq. (\ref{rpa}), has been transformed similarly to the variational functional $L\{w;g\}$ given by eq. (\ref{start}) in the VT based on the lower bound.
Considering that $(\mathcal{M}^*)^2=\overline{\rho}^2g^*$, the self-consistent equation (\ref{sp-w}) reads
\begin{eqnarray}
&&\left.
\frac{1}{N}\left(
\frac{\delta\mathcal{A}\{\overline{\rho};iW\}}{\delta W}
\right)\right|_{W=iW^*}=\frac{\overline{\rho}}{2}g^*\nonumber\\
&&
\frac{\mathcal{L}_{\mathrm{RPA}}}{N}=\frac{\Gamma_2\{\overline{\rho};-W^*\}}{N}
+\frac{\overline{\rho}}{2}\int d{\bf r}\,g^*(r)\beta v(r),\nonumber\\
\label{comp-L}
\end{eqnarray}
where $\Gamma_2\{\overline{\rho};-W^*\}$ is the second Lengendre transform of the functional $\mathcal{A}\{\overline{\rho};-W^*\}$:
\begin{eqnarray}
\frac{\Gamma_2}{N}=
\frac{\mathcal{A}\{\overline{\rho};-W^*\}}{N}+\frac{\overline{\rho}}{2}\int d{\bf r}\,g^*(r)W^*(r).
\label{comp-gamma}
\end{eqnarray}
These expressions (\ref{comp-L}) and (\ref{comp-gamma}) in the DT-FI representation correspond to eqs. (\ref{def_L}) and (\ref{def_gamma}) in the VT based on the lower bound, respectively.
Hence, the treatments of the variational functional $L\{w;g\}$ in the VT is found similar to that of the above functional $\mathcal{L}_{\mathrm{RPA}}$, eq. (\ref{rpa}), in the DT-FI form.

Actually, the maximization of $L\{w;g\}$ in the VT based on the lower bound can be regarded as an aspect of the SP approximation of the $W$-field in the DT-FI representation.
Let us return to eq. (\ref{rpa}) for calculating the second derivative of $\mathcal{L}_{\mathrm{RPA}}$ with respect to $W=W_R+iW_I$ with setting that $\rho^*=\overline{\rho}$:
\begin{eqnarray}
\left.
\frac{\delta^2\mathcal{L}_{\mathrm{RPA}}}{\delta W\delta W}
\right|_{W=iW^*}
=\frac{N\overline{\rho}}
{2\{1-\overline{\rho}W^*({\bf k})\}^2
}\geq 0.
\label{w-second}
\end{eqnarray}
The above positivity of the second derivative specifies the SP configuration of the DT-FI around $W=iW^*$ in the complex functional space as follows:
at the SP field, $\mathcal{L}_{\mathrm{RPA}}$ has either a local minimum along the real field of $W_R$ or a local maximum along the imaginary field of $W_I$.
It is thus found from the expression (\ref{rpa}) that $\mathcal{L}_{\mathrm{RPA}}$ is identified with the variational functional $L\{w;g\}$ when $iW$ is a real function (or $W=iW_I$).
In other words, the SP approximation of the $W$-field supports the VT based on the lower bound [32, 80].

The difference is that the $\mathcal{M}$-field (or the density-density correlation field) in the DT-FI theory can be evaluated to provide the MSA closure relation, opposing the VT based on the lower bound where $g$ is not a variable but is fixed, in principle, at the exact correlation function of the original system as mentioned before.

Furthermore, it is instructive to compare our result, eq. (\ref{rpa}), with the expression of the VT based on the upper bound in the absence of external field, or the WCA perturbation theory for uniform liquids [1, 2, 29-31].
Recalling that
\begin{eqnarray}
&&u^*=\beta v+c^*\nonumber\\
&&w^*=-c^*\nonumber\\
&&g^* u^*=0,
\label{u-w2}
\end{eqnarray}
in addition to the expression (\ref{mf-energy}) with the superscript $*$ denoting the optimization of these interaction potentials, the RPA functional given by eq. (\ref{lower-result}) is rewritten as
 \begin{eqnarray}
&&F\{\beta v\}=F_0\{u^*\}+\Delta F\{w^*\}
\label{rpa-ub}\\
&&\frac{F_0\{u^*\}}{V}=\frac{\overline{\rho}^2}{2}\int dr\,g^*(r)u^*(r)
+\overline{\rho}\ln\overline{\rho}-\overline{\rho}\\
&&\qquad\qquad-\frac{1}{2}\int_{{\bf k}}\ln\left[1+\overline{\rho}h^*({\bf k})\right]\nonumber\\
&&\frac{\Delta F\{w^*\}}{V}
=\frac{\overline{\rho}^2}{2}\int dr\,g^*(r)w^*(r)-\frac{\overline{\rho}}{2}
\label{rpa-comparison}
   \end{eqnarray}
where use has been made of the following relation [1, 6-10]:
\begin{eqnarray}
\ln\left[1+\overline{\rho}h^*({\bf k})\right]
=-\ln\left[1-\overline{\rho}c^*({\bf k})\right].
\end{eqnarray}
The VT based on the upper bound, on the other hand, explores the reference interaction potential $u^*$ that is optimized to minimize the variational free energy functional given by the rhs of the inequality that [1, 2, 21-31]
\begin{eqnarray}
F\{\beta v\}\leq
F_0\{u\}+\frac{\overline{\rho}^2}{2}\int dr\,g_0(r)w(r),
\label{ub}
\end{eqnarray}
where $g_0$ represents radial distribution function of a reference system of particles interacting via the reference interaction potential $u$.
If $g^*$ is close to the optimized function of $g_0$ and we ignore the last term on the rhs of eq. (\ref{rpa-comparison}), it is allowed to regard eq. (\ref{rpa-ub}) as the minimum of the variational free energy functional given by the rhs of eq. (\ref{ub}).
These discussions reveal a connection with the WCA perturbation theory (or the uniform version of the local molecular field theory):
the mean-field perturbative contribution in our DT-FI theory ($\Delta F\{w^*\}$ in eq. (\ref{rpa-comparison})) includes the information on short-range correlations, similarly to the WCA perturbation theory [1, 2, 29-31, 59].

\section{Concluding remarks}

We have thus obtained a set of self-consistent equations that consists of the SP equations regarding one-body and two-body fields as follows: (i) the MMF equation (\ref{rho_mmf3}) of the density field $\rho^*$, (ii) the OZ-like equation (\ref{oz2}) of the interaction potential $W^*$, and (iii) the equivalent of the MSA closure relation (\ref{smsa}) of the radial distribution function $(\mathcal{M}^*)^2=\overline{\rho}^2g^*$.
Compared with the local molecular field equation (\ref{LMF}) [59-62] on which we focus, the advantage is that the MMF equation (\ref{rho_mmf3}) derived from the DT-FI theory specifies the long-ranged interaction potential $w(r)$ in eq. (\ref{LMF}) with minus the direct correlation function $c^*(r)$ given in the MSA, and that we can improve the interaction potential used in the MMF equation (i.e., $w$ in eq. (\ref{LMF})) systematically, starting with the primary approximation performed herein, even though there is a deviation of our short-ranged potential $u=\beta v+c^*$ from that of the local molecular field theory [32] which has been adjusted to mimic the short-range correlations of the true system.

We have also demonstrated that the equation set satisfies the three requirements, (R1) to (R3), described at the end of section II.
It is the first requirement (R1) that guarantees the equivalence of the DT-FI theory and the conventional field theory based on the HS transformation, which has been met by the result that eqs. (\ref{sp-a-rho}) and (\ref{sp-mmf}) obtained from the density functional $\mathcal{A}_{\mathrm{mf}}\{\rho;iW\}$ is of the same form as the mean-field equations (\ref{app_mf}) and (\ref{app_mf2}) given by the HS representation.
We have also confirmed in section V that the DT-FI theory is compatible not only with the HS form [34-43] but also with the VT based on the lower bound [1, 32, 33, 80].
Correspondingly, the second requirement (R2) has been verified by the derivation of eq. (\ref{twobody-solution}).
It is most important that the SP approximation of the $\mathcal{M}$-field in the DT-FI theory has generated the equivalent of the MSA closure relation (\ref{smsa}) satisfying the third requirement (R3).

In terms of the free energy functional, the essential result is represented by eq. (\ref{psi-result}).
It is found from eqs. (\ref{assumption}) to (\ref{lower-result}) that the SP approximations of the DT-FI theory yield, without the input of the LST, the RPA functional of the Helmholtz free energy given by eq. (\ref{lower-result}).
Seemingly, this feature of our formulation is similar not only to the VT based on the lower bound [1, 32, 33, 80] but also to the optimized RPA [24-28];
however, the DT-FI theory requires no reference system as well as the CFT, which is different from the optimized RPA that is relevant especially to the restricted primitive model of electrolytes where the knowledge on hard sphere fluids is input as a reference system [1, 2, 24-28].

In addition, the Ramakrishnan-Youssouf free energy functional in the DFT [11-14, 57] seems to underlie eq. (\ref{rho_mmf3}) and to imply the HNC approximation of the direct correlation function;
however, there are two differences between our results and the Ramakrishnan-Youssouf representation in the DFT.
First, the DT-FI theory verifies that the MSA of the direct correlation function should be used in MMF equation (\ref{rho_mmf3}), unlike the conventional discussions in the DFT.
Second, our results have added the logarithmic term (the second term on the rhs of eq. (\ref{rpa})) to the mean-field free energy functional $\mathcal{A}_{\mathrm{mf}}\{\rho^*;-c^*\}$, being of the same form as the Ramakrishnan-Youssouf free energy functional;
it follows that the MSA has been validated in our formulation.

The relationship between the DT-FI theory and the conventional LST will be made clearer by investigating what approximations should be performed in the field theoretic perturbation around the SP field in order to derive the HNC functional [6-10, 15, 16, 19, 20, 80] of the Helmholtz free energy going beyond the RPA functional, which remains to be addressed along the line of the recent field theoretic formulation in terms of the VT based on the lower bound [80].
Furthermore, we need to clarify the advantage of the DT-FI theory over the LST through the results that are deviated from the LST due to the consideration of higher-order correction terms of $\mathcal{A}\{\rho;iW\}$ including fluctuations of two-body fields ($W$ and $G$).
It is now well established, both theoretically and experimentally, that the dynamical glass transition is accompanied by growing heterogeneity of local correlations, or the critical behavior of four-point correlation function describing the fluctuations of two-point correlations [76-79];
therefore, relevant results could be obtained from investigating fluctuations of two-body fields in glass-forming liquids using the present framework.

\appendix
\section{Derivation of eqs. (\ref{app_mf}) and (\ref{app_mf2})}

To perform the Hubbard-Stratonovich (HS) transformation [34-43] of eq. (\ref{f-start}), we first incorporate the density field $\rho({\bf x})$ into eq. (\ref{f-start}) using the following identity: 
   \begin{eqnarray}
  1&=&\int\,D\rho\,\Delta_N\prod_{\{{\bf x}\}}\>\delta\left[
   \rho({\bf x})-\hat{\rho}({\bf x})
   \right]\nonumber\\
\Delta_N&=&\delta\left\{\int d{\bf x}\rho({\bf x})-N\right\},
\label{rho-identity}
   \end{eqnarray}
where $\Delta_N$ has been added to impose the conservation of the total particle number $N$ on the density field.
Equation (\ref{f-start}) multiplied by the rhs of eq. (\ref{rho-identity}) is
\begin{eqnarray}
  e^{-F\{\beta v\}}&=&\int\,D\rho\,\Delta_N\mathrm{Tr}\prod_{\{{\bf x}\}}\>\delta\left[
   \rho({\bf x})-\hat{\rho}({\bf x})
   \right]\nonumber\\
&&\qquad\times\exp\left[-\beta U\{\rho=\hat{\rho};\beta v\}\right],
\label{rho-replace}
\end{eqnarray}
where the density variable in the interaction energy $U$ is replaced by the c-number field $\rho$.

It follows from the Fourier transform of eq. (\ref{rho-identity}) that a one-body potential field $\psi$, dual to $\rho$, is created:
 \begin{eqnarray}
  \prod_{\{{\bf x}\}}\>\delta\left[
   \rho({\bf x})-\hat{\rho}({\bf x})\right]
=\int\,D\psi\,e^{\int d{\bf x}i\psi({\bf x})
   \left\{\hat{\rho}({\bf x})-\rho({\bf x})\right\}}.
\label{rho-fourier}
   \end{eqnarray}
Correspondingly, eq. (\ref{rho-replace}) is transformed to
\begin{eqnarray}
&&e^{-F\{\beta v\}}=\int D\psi\int D\rho\int d\alpha\,e^{i\alpha\left\{\int d{\bf x}\rho({\bf x})-N\right\}}\nonumber\\
&&\qquad\qquad\qquad\times\mathrm{Tr}\,
e^{\int d{\bf x}i\psi({\bf x})
  \left\{\hat{\rho}({\bf x})-\rho({\bf x})\right\}-\beta U\{\rho;\beta v\}}\nonumber\\
&&=\int D\psi\int D\rho\int d\alpha
\>e^{-\beta\left[U\{\rho;\beta v\}-TS\{\rho;\psi\}\right]+i\alpha\left\{\int d{\bf x}\rho({\bf x})-N\right\}}.\nonumber\\
\label{F-rho-psi}
\end{eqnarray}
In the last line of eq. (\ref{F-rho-psi}), $S$ denotes the entropic contribution given by
\begin{eqnarray}
&&-\frac{S}{k_B}\{\rho;\psi\}=\int d{\bf x}i\psi({\bf x})\rho({\bf x})-\ln\left\{\mathrm{Tr}\,
e^{\int d{\bf x}i\psi({\bf x})\hat{\rho}({\bf x})}\right\}
\nonumber\\
&&=\int d{\bf x}i\psi({\bf x})\rho({\bf x})-N\ln\left\{\frac{\int d{\bf x}e^{i\psi({\bf x})}}{N}\right\}-N-C,
\nonumber\\
\label{s-rho-psi}
\end{eqnarray}
where use has been made of the Stirling's approximation $\ln N!\approx N\ln N-N$ and the constant term $C=\ln\Lambda^{-3N}$ will be omitted for simplicity in the following.
The Gaussian integration over the $\rho$-field in eq. (\ref{F-rho-psi}) yields
\begin{eqnarray}
&&e^{-F\{\beta v\}}=\int D\psi\int d\alpha\,e^{-\mathcal{H}\{\beta v;\psi\}}\nonumber\\
&&\mathcal{H}\{\beta v;\psi\}=\frac{1}{2}\int d{\bf x}\int d{\bf y}\,
\widetilde{\psi}({\bf x})(\beta v)^{-1}(|{\bf x}-{\bf y}|)\widetilde{\psi}({\bf y})\nonumber\\
&&-N\ln\left\{\frac{\left\{\int d{\bf x}e^{i\psi({\bf x})}\right\}}{N}\right\}-N-\frac{N}{2}\beta v(0)+iN\alpha,\nonumber\\
\label{app_hs}
\end{eqnarray}
using the shifted potential, $\widetilde{\psi}=\psi-i\phi-\alpha$.

The SP equation with respect to $\psi$ is
\begin{equation}
\left.
\frac{\delta H\{\beta v;\psi\}}{\delta\psi}\right|_{\psi=i\psi^*}=0,
\label{psi-sp}
\end{equation}
which reads
\begin{eqnarray}
&&\widetilde{\psi}^*({\bf x})
=\int d{\bf y}\,\beta v(|{\bf x}-{\bf y}|)\frac{Ne^{-\psi^*({\bf y})}}{\int d{\bf y}e^{-\psi^*({\bf y})}}\nonumber\\
&&\widetilde{\psi}^*({\bf x})=\psi^*({\bf x})-\phi({\bf x})-\alpha^*,
\label{psi-sp2}
\end{eqnarray}
where a constant $\alpha^*$ arises from the constraint $\Delta_N$.
Equation (\ref{psi-sp2}), which is identical to eq. (\ref{app_mf}), is the mean-field equation in the HS form. 

Let us further introduce a correlation field $G({\bf x},{\bf y})$ into the HS form using the following identity:
    \begin{eqnarray}
  1&=&\int\,DG\,\prod_{\{{\bf x}\}}\prod_{\{{\bf y}\}}\>\delta\left[
  \frac{G({\bf x},{\bf y})}{2}-\frac{G_{\rho}({\bf x},{\bf y})}{2}
   \right],
\label{app_g-identity}
   \end{eqnarray}
whose Fourier transformed representation is given by
\begin{eqnarray}
  1&=&\int\,DW\int\,DG\,\,\nonumber\\
   &&\times\exp\left[\int\int d{\bf x}d{\bf y}\frac{iW({\bf x},{\bf y})}{2}
   \left\{G({\bf x},{\bf y})-G_{\rho}({\bf x},{\bf y})\right\}
   \right].
\nonumber\\
\label{app_g-fourier}
   \end{eqnarray}
Inserting eq. (\ref{app_g-fourier}) into eq. (\ref{F-rho-psi}), the Gaussian integration over the $\rho$-field yields
\begin{eqnarray}
&&e^{-F\{\beta v\}}\nonumber\\
&&=\int DW\int DG\int D\psi\int d\alpha\,
\mathrm{det}^{-1/2}\left[\frac{2\pi}{iW({\bf x},{\bf y})}\right]
\nonumber\\
&&\quad\times
e^{-\mathcal{H}\{iW;\psi\}
-\frac{1}{2}\int\int d{\bf x}d{\bf y}\,G({\bf x},{\bf y})\left\{
\beta v(|{\bf x}-{\bf y}|)-iW({\bf x},{\bf y})\right\}}.
\nonumber\\
\label{psi-start}
\end{eqnarray}
It is to be noted that $\mathcal{H}\{iW;\psi\}$ is of the same form as $\mathcal{H}\{\beta v;\psi\}$.
Since the SP equation with respect to $\psi$ is
\begin{equation}
\left.
\frac{\delta H\{iW;\psi\}}{\delta\psi}\right|_{\psi=i\psi^*}=0,
\label{psi-sp3}
\end{equation}
we have
\begin{eqnarray}
&&\widetilde{\psi}^*({\bf x})
=\int d{\bf y}\,iW(|{\bf x}-{\bf y}|)\frac{Ne^{-\psi^*({\bf y})}}{\int d{\bf y}e^{-\psi^*({\bf y})}}\nonumber\\
&&\widetilde{\psi}^*({\bf x})=\psi^*({\bf x})-\phi({\bf x})-\alpha^*,
\label{psi-sp4}
\end{eqnarray}
similarly to eq. (\ref{psi-sp2}).
At the SP field of $W=iW^*$, the interaction potential $iW$ in eq. (\ref{psi-sp4}) is replaced by $-W^*$.

Equation (\ref{app_mf2}) is thus verified.
Nevertheless, we immediately find a few difficulties in dealing with the above representation (\ref{psi-start}).
First, it is necessary to manipulate the determinant term of $\mathrm{det}^{-1/2}(2\pi/iW)$ when evaluating the FI of the $W$-field.
In addition, $-W^*$ is eventually brought back to the original interaction potential (i.e., $-W^*({\bf x},{\bf y})=\beta v(|{\bf x}-{\bf y}|)$) in the SP approximation of the $G$-field, contradicting the SP equation of $W$, unless we take into account the contributions due to fluctuations of the interaction potential field around $W^*$.

\section{Derivation of eq. (\ref{oz2})}

Equation (\ref{oz2}) is obtained from eq. (\ref{sp-w}).
We then perform the functional differentiation of $\mathcal{L}_{\mathrm{RPA}}$ with respect to $W$ using the expression (\ref{rpa}).
Reminding that the logarithmic correction term in the RPA functional of eq. (\ref{rpa}) is written as
\begin{eqnarray}
&&-\frac{1}{2}\ln\mathrm{det}\left\{
iW({\bf x},{\bf y})+\frac{\delta({\bf x}-{\bf y})}{\rho^*({\bf x})}
\right\}\nonumber\\
&&=\ln\left[
\int D\Delta\rho\,e^{-\frac{1}{2}\int\int d{\bf x}d{\bf y}\,\left\{
iW({\bf x},{\bf y})+\frac{\delta({\bf x}-{\bf y})}{\rho^*({\bf x})}\right\}\Delta\rho\Delta\rho'}
\right],\nonumber\\
\label{det}
\end{eqnarray}
it is found that
\begin{eqnarray}
&&\left.
\frac{\delta}{\delta W}\left[\frac{1}{2}\ln\mathrm{det}\left\{
iW({\bf x},{\bf y})+\frac{\delta({\bf x}-{\bf y})}{\rho^*({\bf x})}
\right\}
\right]\right|_{W=iW^*}
\nonumber\\
&&
\hspace{-5mm}=\frac{i}{2}\left[
\frac
{\int D\Delta\rho\,(\Delta\rho\Delta\rho')\,e^{-\frac{1}{2}\int\int d{\bf x}d{\bf y}\,\left\{
-W^*({\bf x},{\bf y})+\frac{\delta({\bf x}-{\bf y})}{\rho^*({\bf x})}\right\}\Delta\rho\Delta\rho'}}
{\int D\Delta\rho\,e^{-\frac{1}{2}\int\int d{\bf x}d{\bf y}\,\left\{
-W^*({\bf x},{\bf y})+\frac{\delta({\bf x}-{\bf y})}{\rho^*({\bf x})}\right\}\Delta\rho\Delta\rho'}}
\right]\nonumber\\
&&
\hspace{-5mm}=\frac{i}{2}\left<\Delta\rho\Delta\rho'
\right>\nonumber\\
&&
\hspace{-5mm}=\frac{i}{2}\left\{
-W^*({\bf x},{\bf y})+\frac{\delta({\bf x}-{\bf y})}{\rho^*({\bf x})}\right\}^{-1}
\label{det_derivative}
\end{eqnarray}
where $\Delta\rho({\bf x})=\Delta\rho$ and $\Delta\rho({\bf y})=\Delta\rho'$ have been used for brevity, and $\left<\mathcal{O}\right>$ is defined by
\begin{eqnarray}
\left<\mathcal{O}\right>
\equiv
\frac
{\int D\Delta\rho\,\mathcal{O}\,e^{-\frac{1}{2}\int\int d{\bf x}d{\bf y}\,\left\{
-W^*({\bf x},{\bf y})+\frac{\delta({\bf x}-{\bf y})}{\rho^*({\bf x})}\right\}\Delta\rho\Delta\rho'}}
{\int D\Delta\rho\,e^{-\frac{1}{2}\int\int d{\bf x}d{\bf y}\,\left\{
-W^*({\bf x},{\bf y})+\frac{\delta({\bf x}-{\bf y})}{\rho^*({\bf x})}\right\}\Delta\rho\Delta\rho'}}.\nonumber\\
\label{average}
\end{eqnarray}
We also have
\begin{eqnarray}
\left.
\frac{\delta
\mathcal{A}_{\mathrm{mf}}\{\rho^*;iW\}}
{\delta W}
\right|_{W=iW^*}
=\frac{i}{2}G_{\rho^*}.
\label{a_derivative}
\end{eqnarray}
Combining eqs. (\ref{det_derivative}) and (\ref{a_derivative}), the SP equation (\ref{sp-w}) with the expression (\ref{rpa}) is transformed to
\begin{eqnarray}
\mathcal{M}^2=G_{\rho^*}+\left\{
-W^*({\bf x},{\bf y})+\frac{\delta({\bf x}-{\bf y})}{\rho^*({\bf x})}\right\}^{-1}.
\label{m_derivative}
\end{eqnarray}
Introducing the function $k({\bf x},{\bf y})$ as given by eq. (\ref{assumption}), eq. (\ref{m_derivative}) reads
\begin{eqnarray}
&&\rho^*({\bf x})\rho^*({\bf y})k({\bf x},{\bf y})\nonumber\\
&&=-\rho^*({\bf x})\delta({\bf x}-{\bf y})
+\left\{
-W^*({\bf x},{\bf y})+\frac{\delta({\bf x}-{\bf y})}{\rho^*({\bf x})}\right\}^{-1},
\nonumber\\
\label{k_form}
\end{eqnarray}
from which the OZ-like equation (\ref{oz2}) is derived as detailed below.

We make use of the following relation:
\begin{eqnarray}
&&\int d{\bf y}
\,\left\{
-W^*({\bf x},{\bf y})+\frac{\delta({\bf x}-{\bf y})}{\rho^*({\bf x})}\right\}^{-1}
\nonumber\\
&&\qquad\qquad\times
\left\{
-W^*({\bf y},{\bf z})+\frac{\delta({\bf y}-{\bf z})}{\rho^*({\bf y})}\right\}
=\delta({\bf x}-{\bf z}).
\nonumber\\
\label{inverse}
\end{eqnarray}
Performing the same operation on the other terms of eq. (\ref{k_form}), we obtain
\begin{eqnarray}
&&\int d{\bf y}
\,\,\rho^*({\bf x})\rho^*({\bf y})k({\bf x},{\bf y})
\left\{
-W^*({\bf y},{\bf z})+\frac{\delta({\bf y}-{\bf z})}{\rho^*({\bf y})}\right\}\nonumber\\
&&=\rho^*({\bf x})
\left\{-\int d{\bf y}\,
k({\bf x},{\bf y})\rho^*({\bf y})W^*({\bf y},{\bf z})
+k({\bf x},{\bf z})
\right\}
\nonumber\\
\label{k_integral}
\end{eqnarray}
and
\begin{eqnarray}
&&\int d{\bf y}
-\rho^*({\bf x})\delta({\bf x}-{\bf y})
\left\{
-W^*({\bf y},{\bf z})+\frac{\delta({\bf y}-{\bf z})}{\rho^*({\bf y})}\right\}\nonumber\\
&&\qquad\qquad\qquad=\rho^*({\bf x})W^*({\bf x},{\bf z})-\delta({\bf x}-{\bf z}).
\label{rho_integral}
\end{eqnarray}
Going back to eq. (\ref{k_form}), we find that the last term on the rhs of eq. (\ref{rho_integral}) cancels eq. (\ref{inverse}).
Thus it follows from eqs. (\ref{k_form}) to (\ref{rho_integral}) that
\begin{eqnarray}
k({\bf x},{\bf z})=W^*({\bf x},{\bf z})+\int d{\bf y}\,
k({\bf x},{\bf y})\rho^*({\bf y})W^*({\bf y},{\bf z}),\nonumber\\
\label{oz3}
\end{eqnarray}
which is identical to eq. (\ref{oz2}).

\end{document}